\documentclass[prd,preprintnumbers,
twocolumn,
eqsecnum,floatfix,letterpaper,superscriptaddress,nofootinbib]{revtex4}
\usepackage{mathtools}
\usepackage{amsmath,amssymb,graphicx}
\usepackage[printwatermark]{xwatermark}
\usepackage{xcolor}
\usepackage{lipsum}
\usepackage{bm}
\usepackage{color}
\usepackage{booktabs}
\usepackage{tabularx}
\usepackage{times}
\usepackage{microtype}
\usepackage{booktabs}
\usepackage{subfigure}
\usepackage{csquotes}
\usepackage{graphicx}
\usepackage{array}
\usepackage{tikz}
\usetikzlibrary{arrows,positioning} 
\usetikzlibrary{shapes.geometric, arrows}
\newcolumntype{M}[1]{>{\centering\arraybackslash}m{#1}}
\newcolumntype{N}{@{}m{0pt}@{}}

\graphicspath{ {/Users/sayantanidatta/Documents/projects/tgr_svd}}
\usepackage[normalem]{ulem}
\usepackage[varg]{txfonts}
\usepackage{pdfpages}
\usepackage{dirtytalk}
\usepackage[colorlinks, pdfborder={0 0 0}]{hyperref}
\definecolor{LinkColor}{rgb}{0.75 , 0, 0}
\definecolor{CiteColor}{rgb}{0, 0.5, 0.5}
\definecolor{UrlColor}{rgb}{0, 0, 0.75}
\hypersetup{linkcolor=LinkColor}
\hypersetup{citecolor=CiteColor}
\hypersetup{urlcolor=UrlColor}

\allowdisplaybreaks
\def\pcaOne{\delta\hat\phi_{\rm PCA}^{(1)}}
\def\pcaTwo{\delta\hat\phi_{\rm PCA}^{(2)}}

\def\dphithree{\delta\hat\phi_{3}}
\def\dphiFour{\delta\hat\phi_{4}}
\def\dphiFiveEL{\delta\hat\phi_{5l}}

\def\thetaGR{ \vec{\theta}_{{\rm GR}} }
\def\thetaNGR{ \vec{\theta}_{{\rm D}} }

\begin{document}

\title{Parametrized tests of post-Newtonian theory using principal component analysis}
\author{Muhammed Saleem}\email{mcholayi@umn.edu} 
\affiliation{School of Physics and Astronomy, University of Minnesota, Minneapolis, MN 55455, USA}
\affiliation{Chennai Mathematical Institute, Siruseri, 603103, India}
\author{Sayantani Datta}\email{sdatta94@cmi.ac.in} 
\affiliation{Chennai Mathematical Institute, Siruseri, 603103, India}
\author{K. G. Arun}\email{kgarun@cmi.ac.in} 
\affiliation{Chennai Mathematical Institute, Siruseri, 603103, India}
\affiliation{Institute for Gravitation and the Cosmos, Department of Physics, Penn State University, University Park, Pennsylvania 16802, USA}
\author{B. S. Sathyaprakash}  \email{bss25@psu.edu} 
\affiliation{Institute for Gravitation and the Cosmos, Department of Physics, Penn State University, University Park, Pennsylvania 16802, USA}
\affiliation{Department of Astronomy and Astrophysics, Penn State University, University Park, Pennsylvania 16802, USA}
\affiliation{School of Physics and Astronomy, Cardiff University, Cardiff, CF24 3AA, United Kingdom}
\date{\today}

\begin{abstract}

Searching for departures from general relativity (GR) in more than one post-Newtonian (PN) phasing coefficients, called a \emph{multi-parameter test,} is known to be ineffective given the sensitivity of the present generation of gravitational-wave (GW) detectors. {Strong degeneracies in the parameter space} make the outcome of the test uninformative. We argue that Principal Component Analysis (PCA) can remedy this problem by constructing certain linear combinations of the original PN parameters that are better constrained by gravitational wave observations. By analyzing binary black hole events detected during the first and second observing runs (O1 and O2) of LIGO/Virgo, we show that the two dominant principal components can capture the essence of a multi-parameter test.  Combining five binary black hole mergers during O1/O2, we find that the dominant linear combination of the PN coefficients obtained from PCA, $\pcaOne$, is consistent with GR within the $0.38$ standard deviation of the posterior distribution.
{Furthermore, using a set of simulated \emph{non-GR} signals in the three-detector LIGO-Virgo network with designed sensitivities, we find that the method is capable of excluding GR with high confidence as well as recovering the injected values of the non-GR parameters with good precision.} 
\end{abstract}
	
 \maketitle
\section{Introduction}\label{intro}

{Despite the huge success general relativity (GR) has had with the solar system and binary pulsar-based tests,
there has been enormous excitement to test the predictions of the theory in the highly nonlinear regime of the mergers of compact binaries comprising of black holes and neutron stars~\cite{WillLR05,YunesSiemens2013,GairLivRev,Berti:2015itd}. This has been made possible by the recent  detections~\cite{Discovery,GW151226,GW170104,GW170608,GW170814,GWTC1,Venumadhav:2019lyq,Zackay:2019btq,Abbott:2020niy} by the global network of Advanced LIGO~\cite{aLIGO} and the Advanced Virgo~\cite{aVirgo} detectors. Compact binary mergers observed so far are all consistent with the predictions of GR within statistical uncertainties~\cite{O1BBH,O2-TGR,Abbott:2020jks}. However, with the expected enhancement in the sensitivity of LIGO, Virgo, and KAGRA~\cite{KAGRA} detectors in the coming years~\cite{OSD}, we would be in a position to either constrain GR to exceptional precision or detect a deviation from GR. 

One of the most generic tests of GR using GWs, employed on all the GW detections, is the so-called {\it parametrized test of GR}~\cite{AIQS06a, AIQS06b, MAIS10, YunesPretorius09, LiEtal2011, TIGER, YYP2016}. This test searches for potential deviations from GR in the various post-Newtonian (PN) terms in the phase evolution of a signal~\cite{Bliving}.
It is known that the PN phasing coefficients carry the imprints of a variety of physical effects in the general relativistic dynamics of a compact binary. For example, they capture the effects of the `tail' radiation due to the backscattering of the wave by the source's background spacetime~\cite{BS93}, tails of tails~\cite{B98tail,B98quad}, spin-orbit~\cite{K95} and spin-spin interactions~\cite{KWWi93}, among others. In a modified theory of gravity, one might expect one or more of these effects to have behaviors that are qualitatively or quantitatively different from GR~\cite{AlexanderYunes07a,AlexanderYunes07b,CSreview09}, modifying one or more of the PN coefficients.  Hence, precise quantification of the consistency of these coefficients with the predictions of GR is a very powerful test of GR~\cite{AIQS06a,AIQS06b,Chamberlain:2017fjl,Barausse:2016eii}. 

The frequency evolution of the gravitational-wave (GW) phase can schematically be written in the PN approximation as,
\begin{equation}
\Phi(f)=\frac{3}{128\,\eta\,v^5}\sum_{k=0}^N\left[ \phi_k v^k + \phi_{k {\rm l}}\, v^k \ln   v\right],\label{eq:phase}
\end{equation}
where $v=(\pi M f)^{1/3}$ is the PN expansion parameter. $M=m_1+m_2$ is the total mass of the system in the detector frame and $\eta \equiv m_1 m_2/M^2$ is the symmetric mass ratio. $N$ is the highest PN order up to which we currently know the phase evolution. $\phi_k$ and $\phi_{k{\rm l}}$ denote the non-logarithmic and logarithmic PN phasing coefficients, respectively and they depend on the masses and spins of the companion objects, within GR.
In a modified theory of gravity, dependencies of the PN coefficients on masses and spins could be different from GR or they may depend on additional parameters that characterize the new theory. This would potentially deform the unique structure of the PN coefficients in GR. 

The parametrized test searches for such possible deviations from GR by modifying the phasing with dimensionless fractional deviation parameters~\cite{TIGER2012,TIGER2012further,TIGER-2014-Michalis}, 
\begin{equation}
	\phi_a \rightarrow \phi_a^{\rm GR}\left(1+\delta{\hat \phi}_a\right),
	\label{eq:modified}
\end{equation}
where the subscript, `$a$' collectively represents both logarithmic and non-logarithmic coefficients of Eq.~(\ref{eq:phase}). When $\delta{\hat \phi}_a = 0$, it implies ``no deviations'' from GR. GW data is used to quantify the consistency of $\delta{\hat \phi}_a$ with zero which constitutes a ``null' test of GR.

There are different ways in which this null test could be performed. The most general and rigorous approach would be to infer all the eight 
PN deformation parameters \emph {together} from the data~\cite{AIQS06a}. Alternatively, a less general approach could be to test a subset of them simultaneously, assuming the rest to be consistent with GR.   
This class of tests, where a set of two or more of the PN deformation parameters are simultaneously tested, are usually referred to as {\it multi-parameter tests}~\cite{Gupta:2020lxa,Datta:2020vcj,Carl:multiparam}.
However, such tests are not very effective and fail to yield meaningful constraints on the deformation parameters due to high correlations among themselves~\cite{AIQS06a,LiEtal2011,TOG-GW150914}. 

A more pragmatic as well as more restrictive approach is to measure only one deformation parameter at a time while keeping the rest at their GR values~\cite{AIQS06b,LiEtal2011,TOG-GW150914}. 
This is what has been currently employed in the analyses of the LIGO-Virgo data~\cite{GWTC-TGR,Abbott:2020jks}. 
This approach, referred to as {\it one-parameter tests}, leads to eight separate null tests of GR, corresponding to the eight PN coefficients in the phase evolution, although the tests are not necessarily all independent. 
Past studies~\cite{TOG-sampson-2013,TIGER2012} have discussed the efficiency of one-parameter tests to capture generic deviations from GR. 

As the sensitivities of the current-generation GW detectors are not good enough to break the correlations between various parameters, multi-parameter tests are unlikely to be realized shortly. It was recently suggested that multiband observations of a population of stellar-mass black holes may help realize this test in the future~\cite{Gupta:2020lxa,Datta:2020vcj}. Combining the data from stellar-mass binary blackhole coalescences (BBH) in the milli-hertz band of the Laser Interferometer Space Antenna (LISA) and the audio band of ground-based detectors, it would be possible to perform the multi-parameter tests with a precision better than $10\%$~\cite{Gupta:2020lxa,Datta:2020vcj}. However, as multi-band observations of the same systems require the operation of both ground- and space-based detectors, such as LISA, this test may not be possible before the mid-2030s.

Here we propose an alternative, which, in spirit, lies between one-parameter tests and the multi-parameter test, but \emph {does} qualify as a test of the PN structure of the GW phasing formula in GR. The proposal is to use principal component analysis (PCA) to identify the \emph{best-measured} linear combinations of a set of PN deformation parameters.
 The bounds on these linear combinations are arguably more effective in testing GR than the one-parameter tests, as they are sensitive to multiple PN coefficients in the phasing formula and consequently tests the PN structure of the phasing in GR.


Previous works in the literature have pioneered the use of PCA in the context of parameterized tests of GR. 
Ref.~\cite{AP12} demonstrated how the leading eigenvalues and eigenvectors can be used to reduce the effective dimensionality of multi-parameter tests~\cite{AP12,Arun:2013bp}, where they used non-spinning binaries and a parameterization where the PN coefficients themselves were treated as test parameters, following~\cite{AIQS06a}.  
A more recent work~\cite{HanSVD},  focused on the binary neutron star merger GW170817~\cite{GW170817}, 
considered simultaneous estimation of adimensional absolute deviations from GR, for the five PN deformation coefficients between 0PN and 2PN.  They used {\tt TaylorF2} waveform approximant~\cite{CF94} for their analysis. With the PCA of the resulting posterior, they found good agreement with GR for the leading linear combination of the five parameters. In this work, we choose to work with a different set of fractional deformation parameters and focus mainly on BBH systems. We also go further by devising a method to combine the bounds on the PCA parameters from multiple events and also demonstrate their effectiveness in recovering beyond-GR injections.

The remainder of the paper is organized as follows. In Sec.~\ref{sec:SVD-TIGER}, we explain how 
 to derive the new deformation parameters using PCA. The results from the selected GW events are discussed in Sec.~\ref{sec:GWevents}. The role PCA could play in detecting a violation of GR, if present, is discussed in Sec.~\ref{sec:nonGR}. Our conclusions are presented in Sec.~\ref{sec:conclusions}. }

\section{Formalism}\label{sec:SVD-TIGER}

\subsection{Basic concept}

Gravitational waves emitted from a coalescing compact binary system, in the frequency domain, can be schematically written as~\cite{SatDhu91},
\begin{equation}\label{waveform}
{\tilde h}(f)={\cal A}(f) \, e^{i\left[  2\pi f\,t_c-\varphi_c+\Phi(f)  \right]},
\end{equation}
where ${\cal A}(f)$ is the amplitude, $t_c$ and $\varphi_c$ are the time of arrival of the GW signal at the detector and the phase of the signal at that epoch, respectively, and $\Phi (f)$ is the PN phasing that has the schematic form shown in Eq.~(\ref{eq:phase}). 
In GR, the PN coefficients that are currently known are  $\vec{\phi}_a^{{\rm \ GR}} = \{ \phi_0, \phi_2, \phi_3, \phi_4, \phi_{5},  \phi_{5 {\rm l}}, \phi_6, \phi_{6{\rm l}}, \phi_7\}$, up to 3.5PN order. 
In the parametrized test of GR, following the parametrization scheme described in Eq.~(\ref{eq:modified}), this would constitute to a set of nine deformation parameters\footnote{The non-logarithmic deformation parameter at 2.5PN, $\delta\hat{\phi}_5$ is not considered as it can simply be absorbed into a redefinition of $\varphi_c$~\cite{B95,AISS05}. },
\begin{equation}
\thetaNGR = \{ \delta\hat{\phi}_0, \delta\hat{\phi}_2, \delta\hat{\phi}_3, \delta\hat{\phi}_4,  \delta\hat{\phi}_{5 {\rm l}}, \delta\hat{\phi}_6, \delta\hat{\phi}_{6{\rm l}}, \delta\hat{\phi}_7\}.
\label{eq:nongr}
\end{equation}

The GR phasing and amplitude terms already depend on several binary parameters that include component masses, spins, luminosity distance, and the angular parameters describing the sky location and orientation of the source. For a precessing binary black hole on quasi-circular orbits, this would count up to 15 parameters. In general, there are \textit{m} GR parameters and \textit{n} non-GR parameters. 
In a \textit{multiparameter} test of GR, this implies to a Bayesian inference problem with an \textit{m+n}-dimensional parameter space, 
\begin{equation}\label{PE}
P(\{ \thetaGR, \thetaNGR \}\,|\,{\mathcal H},d) = \frac{P(\{ \thetaGR,\thetaNGR\} \, | \, {\mathcal H}) P(d\,|\,{\mathcal H},\{\thetaGR, \thetaNGR \})}{P(d\,|\,{\mathcal H})}.
\end{equation}
See  Appendix \ref{basics-Bayesian} for a quick review of Bayesian inference.
Since we are interested in the deformation parameters, we marginalize over the GR parameters,
\begin{equation}\label{eq:marg}
P(\thetaNGR\,|\,{\mathcal H},d) = \int P(\{ \thetaGR, \thetaNGR \}\,|\,{\mathcal H},d) \, d\thetaGR.
\end{equation}
which leaves us with the \textit{n}-dimensional posteriors of $\thetaNGR$. 

As mentioned earlier, in general, these parameters are correlated. Equivalently, the posterior on $\thetaNGR$ has an associated covariance matrix that is non-diagonal. Our proposal is to use the PCA technique to perform a rotation of the parameters, $\thetaNGR$, to an orthogonal basis in which the covariance matrix is diagonal.
 
\begin{figure*}[t]
	\centering
	\includegraphics[scale=0.5]{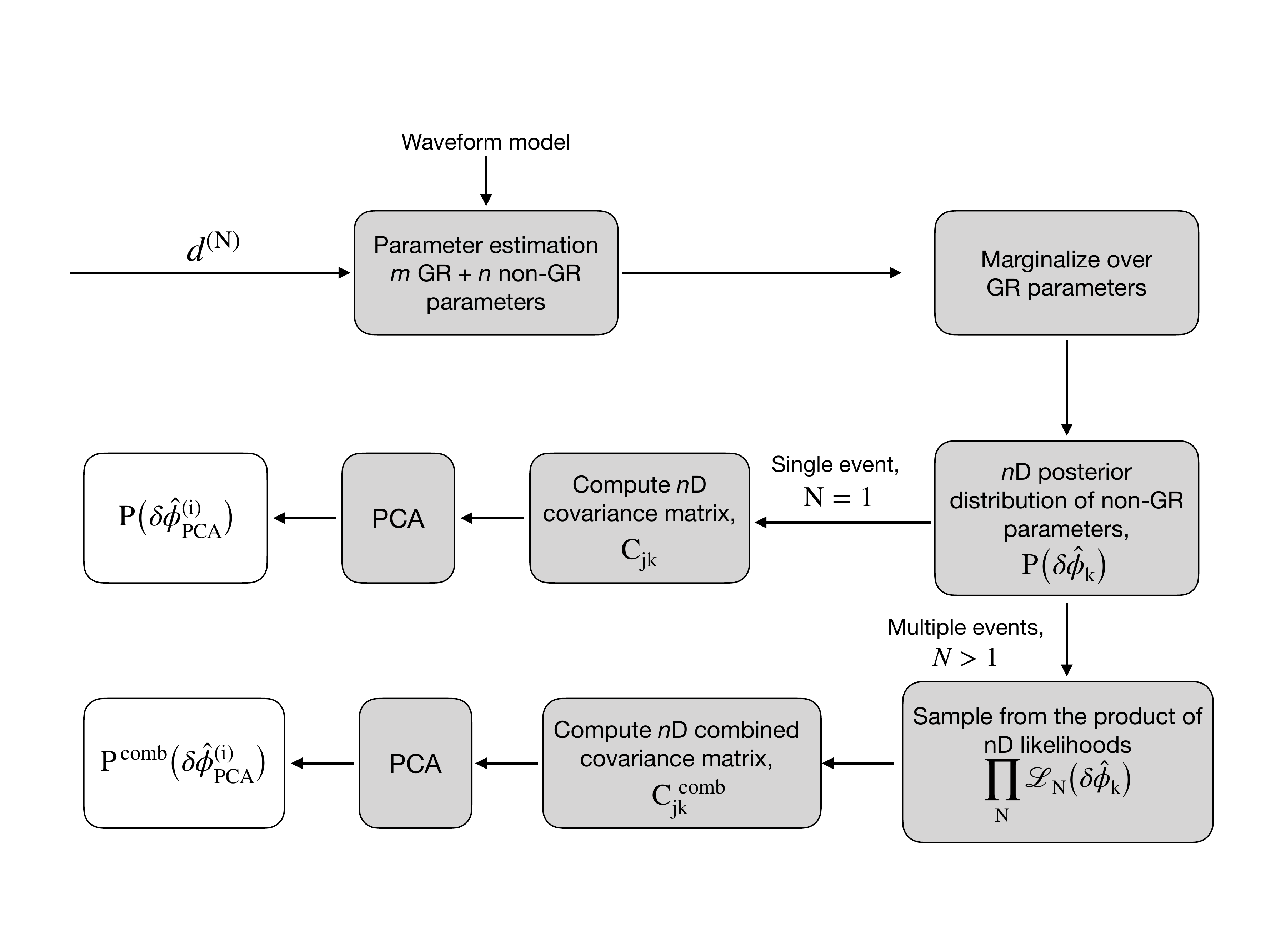}
\label{fig:flowchart}
\caption{Flow chart showing the hierarchy of steps followed to obtain the posterior distribution of PCA parameters, starting from the data, for $n$ number of events. See the text for more details.}
\end{figure*}

In the subsection below, we detail the steps that are followed to construct the new deformation parameters. 
A flowchart of the same has been shown in Fig.~\ref{fig:flowchart}.

\subsection{Construction of new deformation parameters}\label{method-svd}
First, after marginalizing over the GR parameters, we compute the  variance-covariance matrix $\mathbf{C}$ of the $n$-dimensional posterior of the deformation parameters for each event,
\begin{equation}\label{eq:cov}
	C_{jk} = \left\langle   \left({\delta\hat{\phi}_j} - \langle {\delta\hat{\phi}_j} \rangle \right)    \left({\delta\hat{\phi}_k} - \langle {\delta\hat{\phi}_k} \rangle \right)	\right\rangle,
\end{equation}
where, $\delta\hat{\phi}_j$ and $\delta\hat{\phi}_k$ are the respective marginalized posterior samples and the symbol $\left <x\right >$ refers to the expectation value of the random variable $x$. 
We then diagonalize the marginalized covariance matrix and compute the corresponding eigenvalues and eigenvectors 
\begin{equation}\label{svd_decomposition}
C = U S U^T,
\end{equation}
where $S$ is the eigen value matrix which is diagonal and $U$ is a unitary matrix whose columns are the eigenvectors corresponding to each diagonal entry in $S$. 
Algebraically, $U$ represents a transformation of the basis in which the covariance matrix, $C$ is non-diagonal, to a new basis in which the covariance matrix $S$ is diagonal (see Appendix \ref{reconstruction} for more details). 

The relative information carried by each of the eigenvectors can be estimated by looking at the hierarchy of the ratios of the eigenvalues to the smallest one. In our case, the \textit{smallest eigenvalue} is synonymous with the \textit{most informative} eigenvector as it has the smallest error bar. On the other hand, those with very large error bars are least informative, and often their posteriors will resemble the priors themselves and it would be safe to truncate the parameter space by excluding them. We truncate the diagonal covariance matrix by excluding those eigenvalues that are greater than 1000 times the leading eigenvalue\footnote{{This is a choice that is suitable for our purposes, this may have to be revisited in the context of future detectors.}}.  The eigenvectors corresponding to the surviving eigenvalues are the new deformation parameters, 
\begin{equation}
\delta\hat{\phi}_{PCA}^{(i)}=\sum_{k} \alpha^{ik}\,\delta\hat{\phi}_k,
\label{eq:newparam}
\end{equation}
which will be referred to as the  \textit{PCA parameters}. The index $k$ is summed over the number of independent deformation parameters ($n$) in the original multi-parameter test
and the index $i$ is the number representing the PCA parameter in the order of their dominance (for instance, $i=1$ for the most dominant one). The coefficients $\alpha$ are the components of the matrix $U$. 
Hence, Eq.~(\ref{eq:newparam}) defines the new deformation parameters that are linear combinations of all the original PN deformation parameters and the measurement of each of the new parameters carries the essence of \emph{multi-parameter tests}. 

\subsection{Combining information from multiple events}\label{sec:combining}
One can usually combine the bounds from multiple events by multiplying the marginalized likelihoods of every single event. This implicitly assumes that the true value of the deformation parameters is the same across the events. Despite this assumption, this approach is possible only for combining the original deformation parameters, not for the PCA parameters. This is because, PCA parameters, as defined in Eq.~(\ref{eq:newparam}), are unique linear combinations for each event with the coefficients $\alpha^{ij}$ being functions of the source parameters.

We follow the following approach to overcome this. First, we compute the \textit{n}-dimensional marginalized likelihood for the original deformation parameters, $\thetaNGR$, for all the events following Eqs.~(\ref{PE}) and~(\ref{eq:marg}). The combined posterior samples are then obtained by sampling from the product of all the individual \textit{n}-dimensional likelihoods. Practically, we achieve this with the help of Gaussian kernel density estimates constructed for the 
\textit{n}-dimensional likelihoods of each event and taking their product. The PCA parameters are then computed for the combined \textit{n}-dimensional posterior by diagonalizing as described in sec.~\ref{method-svd}.

%
\section{Results and discussion}\label{sec:GWevents}

Having introduced the method, we now discuss its application on GW events, both real events and simulated events. Limited by the current sensitivity of the GW detectors,  we consider six dimensional multi-parameter test with deformation parameters introduced at  orders from 1.5PN through 3.5PN. This implies, $\thetaNGR = \{\delta\hat\phi_{3}, \delta\hat\phi_{4}, \delta\hat\phi_{5l}, \delta\hat\phi_{6}, \delta\hat\phi_{6l}, \delta\hat\phi_{7} \}$ which implicitly assumes that the leading order PN deformations ($\delta\hat\phi_{0}, \delta\hat\phi_{1},\delta\hat\phi_{2}$) are consistent with GR.~\footnote{The consistency of lower PN coefficients with GR may be naturally expected in effective field-theoretic extension of GR where modifications may appear at relatively higher PN orders~\cite{Endlich:2017tqa,Sennett:2019bpc}.} Together with the 15 GR parameters, this leads to a  21-dimensional parameter space. We use the {\tt LALInference} package~\cite{lalinference} and the  {\tt IMRPhenomPv2}~\cite{Khan2016} waveform model to perform the Bayesian inference described in Eqs.~(\ref{PE}) and~(\ref{eq:marg}). {The inspiral part of the {\tt IMRPhenomPv2} waveform is deformed as shown in Eq.~\ref{eq:modified} to model the possible deviations~\cite{TOG-GW150914}.}

For every single event, we first obtain the six-dimensional posterior of $\thetaNGR$ and then use Eqs.~(\ref{eq:cov}),~(\ref{svd_decomposition}) and~(\ref{eq:newparam}) to compute the posterior samples for the PCA parameters. Following the truncation criterion suggested in sec.~\ref{method-svd}, we found that only the leading two PCA parameters survive the criterion. Therefore, we show the results only for the two leading PCA parameters $\pcaOne$ and $\pcaTwo$, arguing that they are good enough to reconstruct the likelihood to a good approximation, for the tolerance set by our truncation criterion.   

\subsection{Application to selected BBHs detected during O1/O2}\label{sec:SVDO1O2}
We first discuss the results obtained by applying the method on the events detected during the first two observing runs O1 and O2 of the Advanced LIGO and Advanced Virgo detectors.  Specifically, we choose BBH mergers GW150914, GW151226 from O1 and  GW170104, GW170608, GW170814 from O2. For each of these events, the signal-to-noise ratio in the inspiral phase is higher than 6 and they are also the events for which the parameterized tests in Refs.~\cite{O1BBH,O2-TGR} were performed. We have not performed the analysis on the binary neutron star merger GW170817 mainly because it is computationally expensive due to a large number of signal cycles in the sensitivity band of LIGO and Virgo (see \cite{HanSVD} for a detailed discussion in the case of GW170817). 
\begin{figure}
\centering
\begin{subfigure}
  \centering
  \includegraphics[width=0.4\textwidth]{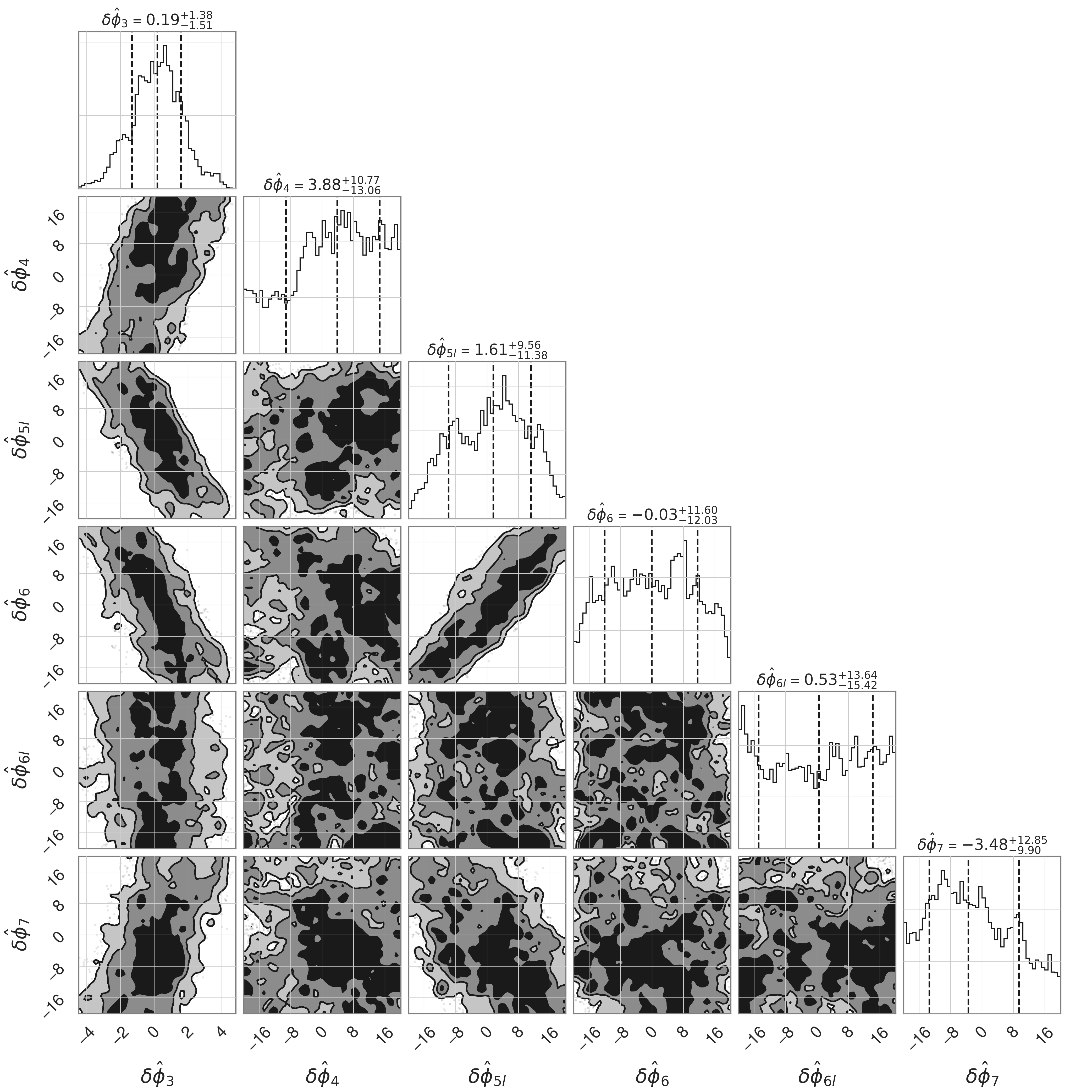}
\end{subfigure}
\begin{subfigure}
  \centering
  \includegraphics[width=0.4\textwidth]{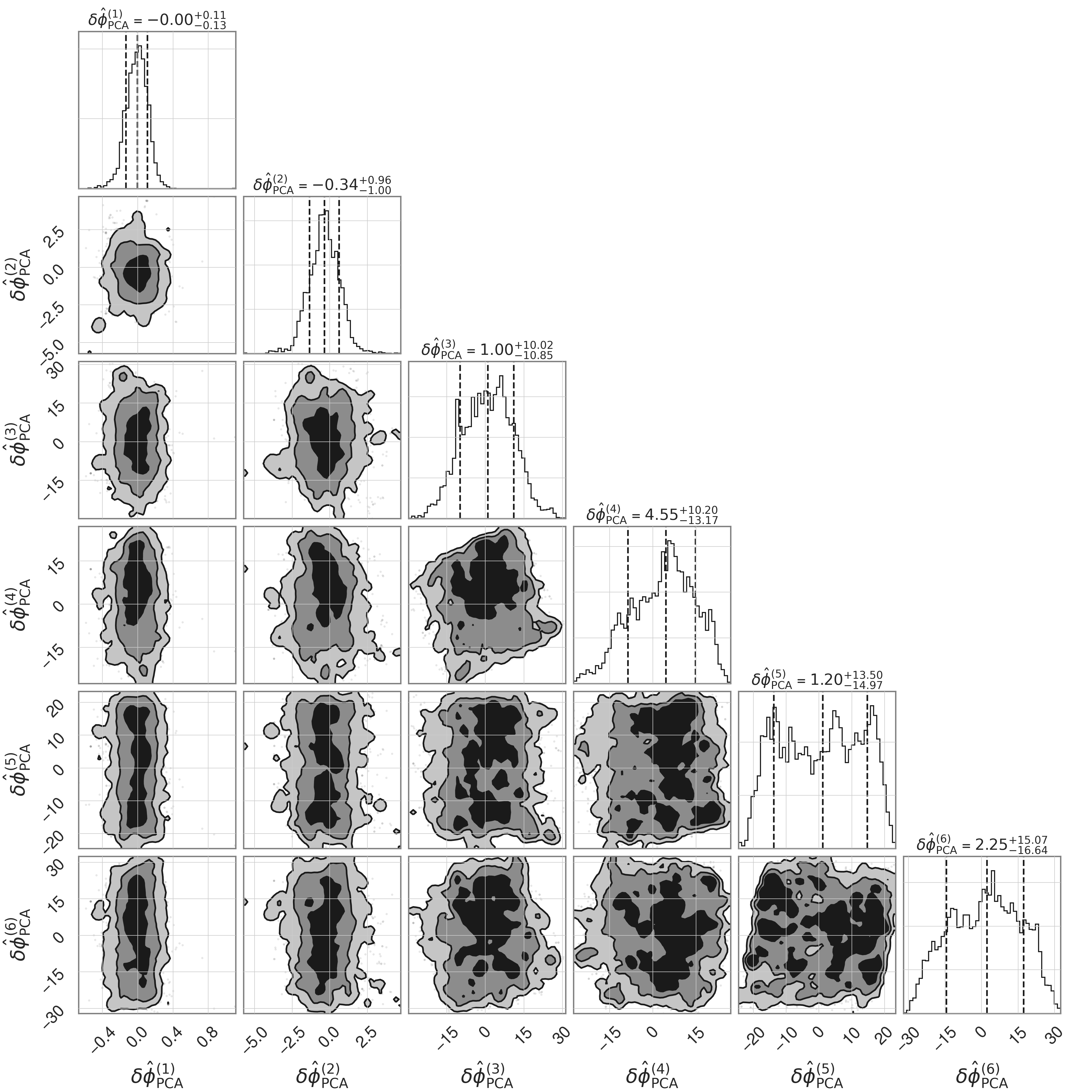}
\end{subfigure}
\caption{Six dimensional corner plots showing the bounds and covariances among the different parameters from the \textit{multi-parameter} analysis of GW151226 before (top) and after (bottom) performing PCA. Deformation of the PN phasing starts at 1.5PN order thus six independent deformation parameters are simultaneously estimated from the data. The new linear combinations are obtained by the PCA of the six dimensional posteriors after marginalizing over the GR parameters. It is evident that the use of PCA dramatically improves the efficiency of the null test. Eigenvectors corresponding to larger eigenvalues (subdominant new parameters) show noisy features in the marginalized 1D posteriors as the dominant parameters contribute the most to the likelihood, making the subdominant new parameter posteriors noisier.}
\label{fig:GW151226-real}
\end{figure}

To understand the results, we first focus on GW151226. The results are shown in the corner plots in Fig.~\ref{fig:GW151226-real}. The top plot shows the bounds and covariances of the original deformation parameters  $\{\delta\hat\phi_{i}\}$ while the bottom plot shows the same for the PCA deformation parameters. 
As expected, by virtue of PCA, the widths of the first two dominant linear combinations are much smaller than those of the original parameters. For the subdominant linear combinations, the widths are as bad or worse than the original parameters, since most of the information is already captured by the leading new parameters. 
The shapes of the contours underscore the fact that correlations among the new set of parameters are, as expected, mostly removed by the PCA thereby bringing the covariance matrix to a diagonal form. The posteriors of the least dominant linear combinations show multi-modal features. This is naturally expected as most of the information in the data goes into the construction of the leading PCA parameters, making the sub-dominant ones noise-dominated.

Quantitatively, the two leading PCA parameters are constrained as  $\pcaOne = 0.00^{+0.18}_{-0.21}$ and $\pcaTwo = -0.34^{+1.55}_{-1.65}$ where the numbers shown are the median values and the 90\% credible error-bars. The GR values, \textit{i.e.,} $\pcaOne = 0.0$ and $\pcaTwo = 0.0$ are recovered at just $0.03\sigma$ and $0.34\sigma$ away from the medians respectively, where $\sigma$ is the standard deviation of the respective posterior distributions. 

Table~\ref{tab:realevents-pca} summarizes the results from all the events. Apart from $\pcaOne$ and $\pcaTwo$, for comparison, the table also shows the bounds on the deformation parameters at the 1.5PN, 2PN and the logarithmic term at 2.5PN order from the one-parameter tests as reported in  \cite{O1BBH}. 
Their selection for the comparison follows from the fact that they are the best-measured deformation parameters from one-parameter tests starting from 1.5PN. For all the five events, we note that the bounds on $\pcaOne$ are comparable to the bounds on the 1.5PN deformation parameter from the one-parameter tests.
%
 \begin{table*}
 	\begin{center}
 		\begin{tabular}{ |M{1.8cm} | M{1.5cm} | M{1cm} | M{1.5cm} | M{1cm} | M{1.5cm} | M{1cm} | M{1.5cm} | M{1cm}| M{1.5cm} | M{1cm} |}
 			\hline
			\multicolumn{1}{| c |}{} &  \multicolumn{4}{ c |}{}  &  \multicolumn{6}{ c |}{}  \\[-6pt]
			\multicolumn{1}{| c |}{} &  \multicolumn{4}{ c |}{\textbf{  PCA deformation parameters}}  &  \multicolumn{6}{ c |}{\textbf{ TIGER one-parameter deformations }} \\[4pt]
			\cline{2-11}  
			\multicolumn{1}{| c |}{}  & \multicolumn{2}{ c |}{} &  \multicolumn{2}{ c |}{} &  \multicolumn{2}{ c |}{}   &  \multicolumn{2}{ c |}{} 	 & \multicolumn{2}{ c |}{} \\[-6pt]
			\multicolumn{1}{| c |}{\textbf{Event}}  & \multicolumn{2}{ c |}{\textbf{ $\pcaOne$}} &  \multicolumn{2}{ c |}{\textbf{ $\pcaTwo$}} &  \multicolumn{2}{ c |}{\textbf{ $\delta\hat{\phi_3}$}}  &  \multicolumn{2}{ c |}{\textbf{ $\delta\hat{\phi_4}$}} 	 & \multicolumn{2}{ c |}{\textbf{ $\delta\hat{\phi_{5l}}$}} \\[2pt]
			\cline{2-11}  

 			& Median \&  & GR & Median \&  & GR& Median \&  & GR& Median \&  & GR& Median \&  & GR  \\[2pt]
 			& 90\% errors & value at & 90\% errors & value at & 90\% errors & value at & 90\% errors & value at & 90\% errors & value at   \\[2pt]
 			\hline\hline
 			
 			&&&&&&&&&&\\[-4pt]		
 GW150914  & $-0.22^{+0.19}_{-0.16}$   &  2.05$\sigma$  &  $3.58^{+4.85}_{-4.37}$   &  1.26$\sigma$  &  $0.22^{+0.2}_{-0.2}$   &  1.79$\sigma$         & $-1.92^{+1.7}_{-1.63}$     &  1.91$\sigma$  & $0.7^{+0.55}_{-0.58}$      &  2.01$\sigma$ \\[5pt]
 GW151226  & $0.0^{+0.18}_{-0.21}$     &  0.03$\sigma$  & $-0.34^{+1.55}_{-1.65}$   &  0.34$\sigma$ & $-0.01^{+0.15}_{-0.2}$   &  0.1$\sigma$         & $0.07^{+1.62}_{-1.33}$     &  0.08$\sigma$   & $-0.03^{+0.48}_{-0.67}$   &  0.1$\sigma$ \\[5pt] 
 GW170104  & $0.56^{+0.57}_{-0.51}$    &  1.68$\sigma$    & $4.44^{+14.06}_{-15.3}$  &  0.48$\sigma$  & $-0.48^{+0.42}_{-0.65}$   &  1.43$\sigma$       & $3.73^{+4.62}_{-3.56}$     &  1.47$\sigma$   & $-1.41^{+1.26}_{-1.61}$   &  1.56$\sigma$ \\[5pt]
 GW170608  & $-0.04^{+0.15}_{-0.2}$     &  0.35$\sigma$  & $-0.03^{+1.27}_{-1.05}$   &  0.04$\sigma$ & $0.05^{+0.12}_{-0.13}$   &  0.63$\sigma$        & $-0.26^{+1.07}_{-1.09}$    &  0.4$\sigma$    & $0.09^{+0.42}_{-0.42}$    &  0.36$\sigma$ \\[5pt]
 GW170814  & $-0.14^{+0.21}_{-0.19}$   &  1.14$\sigma$  & $-2.88^{+2.18}_{-2.13}$   &  2.17$\sigma$ & $0.07^{+0.21}_{-0.22}$   &  0.53$\sigma$        & $-0.45^{+1.71}_{-1.64}$    &  0.44$\sigma$  & $0.1^{+0.69}_{-0.63}$      &  0.25$\sigma$ \\[5pt] 
 Combined   & $-0.02^{+0.07}_{-0.08}$   &  0.38$\sigma$   & $-0.39^{+0.57}_{-0.76}$  &  0.96$\sigma$ & $0.05^{+0.07}_{-0.09}$   &  0.96 $\sigma$     & $-0.36^{+0.7}_{-0.69}$      &  0.86$\sigma$   & $0.14^{+0.23}_{-0.26}$   &  0.95$\sigma$ \\[3pt]
\hline
\end{tabular}
\end{center}
\caption{The posterior properties of the leading two deformation parameters obtained from PCA. The first five rows show the bounds on the selected five events from GWTC-1 and the sixth row shows the combined bounds from all the five events. }
\label{tab:realevents-pca}
\end{table*}	
As can be noted from Table~\ref{tab:realevents-pca}, the PCA posteriors of GW150914 and GW170104 recover GR values slightly outside the 90\% credible levels. This feature has also been observed in the one-parameter tests. As is evident from the Table~\ref{tab:realevents-pca}, for both  GW150914 and GW170104,  the posteriors of $\dphithree$, $\dphiFour$ and $\dphiFiveEL$ recovered GR just outside their 90\% credible error bars. This feature has been studied and these offsets are very likely due to noise artefacts~\cite{Discovery,GW170104}.
For the other three events, $\pcaOne$, as well as the three aforementioned one-parameter tests, could recover GR with  90\% credibility. 

\subsection{Combined bounds from O1/O2 events}
\begin{figure}[t]
	\centering
	\includegraphics[width=0.4\textwidth]{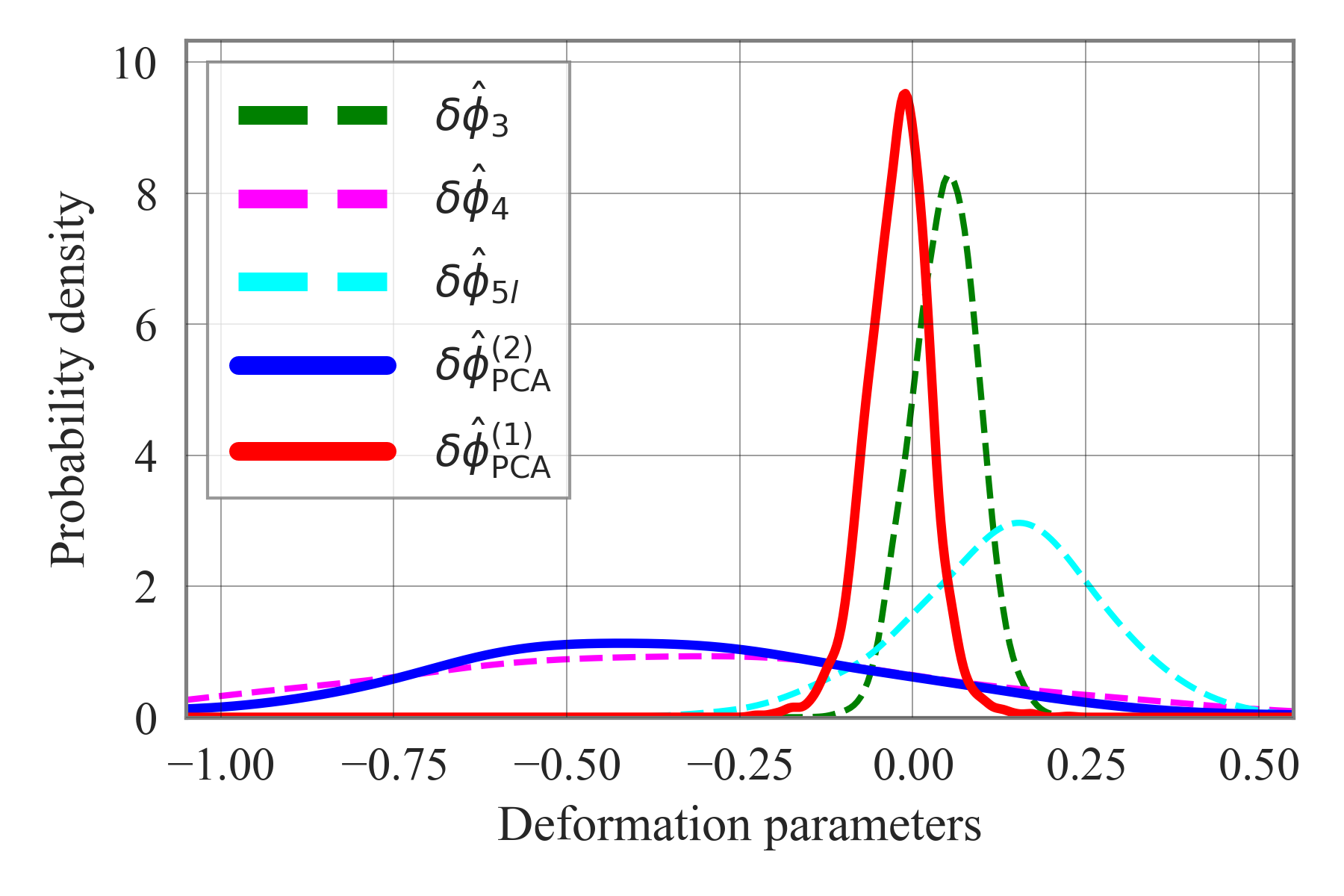}
	\caption{Marginalized combined posterior probability distributions  of five O1/O2 events; GW150914, GW151226, GW170104, GW170608, and GW170814. 
		The solid lines show the two dominant linear combinations obtained from PCA, namely $\pcaOne$ and $\pcaTwo$. They are computed from the combined multi-dimensional posteriors, from all five events. 
		The dashed lines are the combined posteriors on the TIGER deformation parameters $\{\delta\hat\phi_{3}, \delta\hat\phi_{4}$ and $\delta\hat\phi_{5l}$ obtained from one-parameter tests, as reported in \cite{GWTC-TGR}.}
	\label{fig:hist-combined-allfive}
\end{figure}

Fig.~\ref{fig:hist-combined-allfive} shows the combined bounds from all the five events, using the method described in sec.~\ref{sec:combining}.  The posterior probability distributions denoted with solid lines correspond to the two leading PCA parameters obtained from the combined data which naturally are the best estimated linear combinations. These are compared against the combined posterior probability distributions of $\dphithree$, $\dphiFour$ and $\dphiFiveEL$ from one-parameter tests as listed in Table~\ref{tab:realevents-pca}. 

The joint bound on the leading PCA parameter (also shown in the last row of Table~\ref{tab:realevents-pca}), $\pcaOne$ is estimated to $-0.02^{+0.07}_{-0.08}$ at 90\% credibility and the same for $\delta\hat{\phi}_3$, the best-constrained deformation parameter from among the one-parameter tests, is estimated to be $0.05^{+0.07}_{-0.09}$. Even though these are quite comparable, $\pcaOne$ is more consistent with GR in the sense that the GR value (zero) is recovered just 0.38$\sigma$ away from the median whereas the  $\delta\hat{\phi}_3$ recovers GR at 0.96$\sigma$, farther away from the median of the posterior. The sub-leading PCA deformation parameter, $\pcaTwo$ is poorly constrained and peaks away from zero as compared to  $\pcaOne$ which could be an indication that most of the information has already been captured by the leading parameter, making the sub-leading one noisier. The better consistency of the peak of the leading parameter with the GR value is a salient feature. There could be cases where one-parameter tests might yield well-constrained posteriors of deformation parameters, but might peak away from zero indicating a deviation from GR. Besides a genuine GR violation, noise artifacts can also cause such features. We find, based on the analyses of a limited number of events, that similar offsets are seen in the PCA-based parameters as well. The ability of the PCA-based method to mitigate such artifacts will require a more detailed study using noisy injections of GR signals which we reserve for a future publication.

\section{Detecting GR violations}\label{sec:nonGR}

We have seen the efficiency of the PCA-based method to constrain possible deviations from GR. In this section, we explore the capability of this method to detect a deviation from GR, \textit{i.e,}  how well this method can rule out GR if the true signal is non-GR. 

We perform the analysis on a set of 10 non-GR injections in the three-detector LIGO-Virgo network with all of them assumed to be at their respective designed sensitivities. The masses and spins of the injections were randomly drawn from the population models inferred from the O1/O2 detected GW events~\cite{LIGOScientific:2018jsj}.   
For each injection, we introduced fractional deviations from GR at all orders starting from 1.5PN up to 3.5PN by the same amount. For a given injection, this fractional value is chosen by randomly drawing from a normal distribution centered at 0.5 with a standard deviation of 0.1 (See Table~\ref{tab:nongr-pca} for details of the injections). These choices of non-GR parameters are purely arbitrary and the aim here is to assess the ability of the method to detect a GR violation.

\begin{table*}
	\begin{center}
		\begin{tabular}{ |M{1cm} | M{1.9cm} | M{1.5cm} | M{1.5cm} | M{1.5cm} | M{1.5cm} | M{1.9cm} | M{1.7cm} | M{1.7cm} |}
			\hline
			\multicolumn{1}{| c |}{}  & \multicolumn{5}{ c |}{\textbf{ Properties of the injections}} & \multicolumn{3}{ c |}{\textbf{ Properties of  $\pcaOne$}} \\
			\multicolumn{1}{| c |}{}  & \multicolumn{5}{ c |}{} & \multicolumn{3}{ c |}{} \\
			
			\cline{2-9}  
			\textbf{Event} & Component  Masses ($M_\odot$)  & Spins  (aligned)   & $D_L$  (Mpc)  & $\delta\hat{\phi}_{k}$: $k\in$ {\scriptsize \{3,4,5l,6,6l,7\} }& Network  SNR  & Median \&   90\% errors & GR value recovered at & True value recovered at \\ 
			\hline\hline
			&&&&&&&&\\[-4pt]
			Sim 1 &	17.5, 14.2  & 0.2, 0.2  & 801	&  0.65	&  18   & $-0.65^{+0.09}_{-0.09}$  & 10.80$\sigma$ & 0.55$\sigma$ \\[4pt] 
			Sim 2 &	15.7, 12.9  & 0.4, 0.1  & 1891 &  0.64 &  17  & $-0.65^{+0.41}_{-0.25}$  & 2.73$\sigma$   & 0.24$\sigma$ \\[4pt] 
			Sim 3 &	29.0, 28.3  & 0.3, 0.1  & 706	&  0.53	&  54   & $-0.54^{+0.05}_{-0.07}$  & 13.69$\sigma$ & 0.23$\sigma$ \\[4pt]
			Sim 4 &	5.1, 5.1      & 0.3, 0.2  & 847	  &  0.45 &  14   & $-0.26^{+0.25}_{-0.24}$  & 1.74$\sigma$  & 1.33$\sigma$  \\[4pt]
			Sim 5 &	5.4, 5.1     & 0.6, 0.4  & 518	 &  0.68  &  14   & $-0.55^{+0.14}_{-0.11}$  & 6.95$\sigma$  & 1.04$\sigma$ \\[4pt]
			Sim 6 &	17.8, 13.8  & 0.4, 0.3  & 1085 &  0.38 &  27  & $0.36^{+0.13}_{-0.14}$   & 3.82$\sigma$  & 0.48$\sigma$  \\[4pt]
			Sim 7 &	12.7, 9.4   & 0.4, 0.3  & 1063	&  0.41	&  12   & $0.36^{+0.16}_{-0.19}$   & 3.11$\sigma$  & 0.72$\sigma$ \\[4pt]
			Sim 8 &	6.2, 6.1     & 0.5, 0.3  & 823	 &  0.46  &  19   & $0.32^{+0.15}_{-0.29}$   & 2.16$\sigma$  & 1.22$\sigma$ \\[4pt]
			Sim 9 &	19.2, 14.5 & 0.3, 0.1  & 1998 &  0.58  &  19  & $0.52^{+0.23}_{-0.26}$   & 3.13$\sigma$  & 0.59$\sigma$ \\[4pt]
			Sim 10&	5.2, 5.1    & 0.4, 0.4   & 675	&  0.53  &  21   & $0.39^{+0.16}_{-0.23}$   & 3.02$\sigma$   & 1.09$\sigma$ \\[3pt]
			\hline
		\end{tabular}
	\end{center}
\caption{Demonstrating the ability of PCA to detect deviations from GR. For a set of non-GR injections, the table shows the bounds on the leading PCA deformation parameter $\pcaOne$. The GR exclusion by $\pcaOne$ is shown as how many standard deviations away is the GR value (zero) from the median of the posterior. The inclusion of the true value is also quantified in the same manner. It is found that the GR values are excluded at $> 3\sigma$ for most of them. Similarly, the true values of $\pcaOne$ are included within their 1$\sigma$ credible interval for most of the cases. }
	\label{tab:nongr-pca}
\end{table*}	

As explained in sec.~\ref{method-svd}, we first perform the parameter estimation ( by considering zero-noise realization) on the simulated non-GR signals with the aforementioned six free PN deformation parameters. Principal Component Analysis is then performed on the marginalized 6-dimensional posteriors to deduce the dominant linear combinations of the PN deformation parameters that are best estimated. 

\begin{figure*}[t]
	\centering
	\includegraphics[width = 0.9\textwidth]{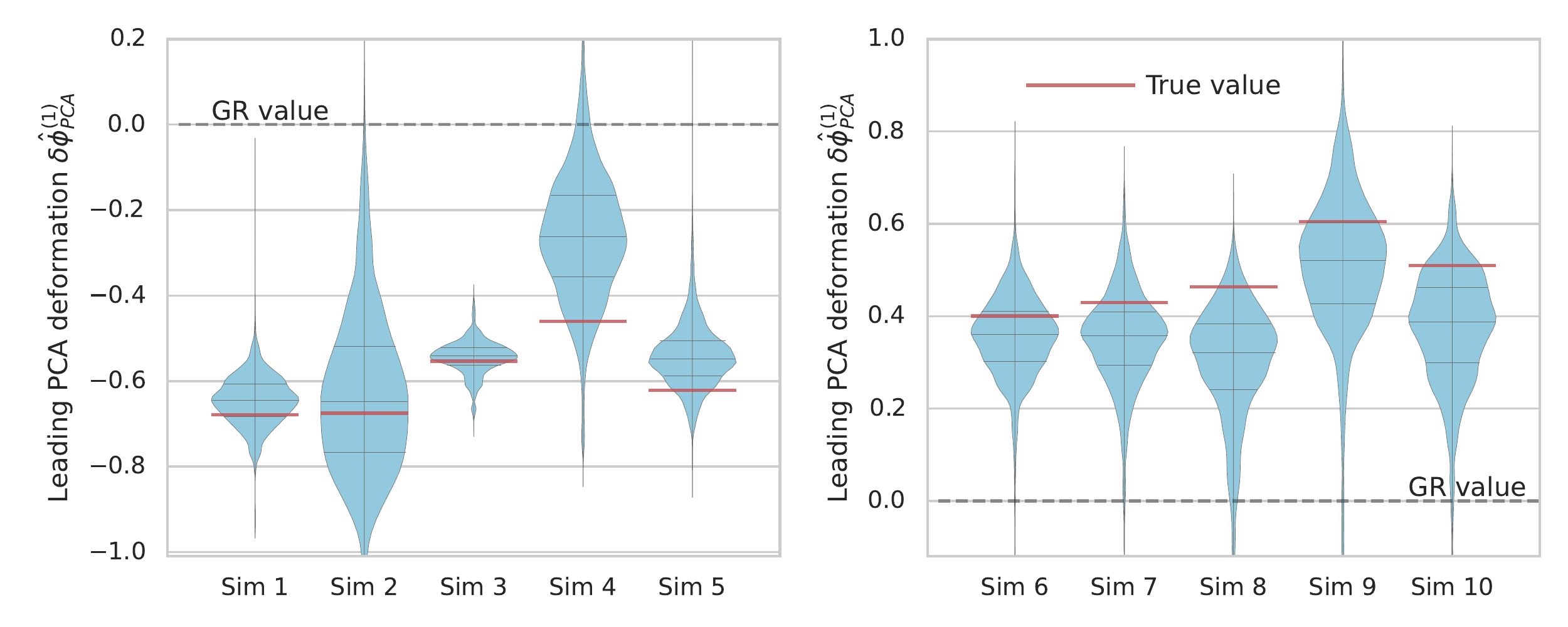}
	\caption{Violin plots showing the recovery of the ten non-GR injections using the PCA of multi-parameter tests. The solid red lines show the injected values and the GR value (zero) is marked by dashed grey lines. The grey solid horizontal lines within the violin plots mark the median and $1\sigma$ bounds. }
	\label{fig:violin-nongr}
\end{figure*}

Fig.~\ref{fig:violin-nongr} shows the violin plots for the posteriors of the leading PCA parameter $\pcaOne$ from the analyses and the quantitative details from them are presented in Table~\ref{tab:nongr-pca}. As is evident, the posteriors are well constrained, with the 90\% credible widths being 0.12 and 0.66, respectively, for the most and the least constrained injections. As for  quantifying the GR exclusion, for each injection, we compute how many standard deviations ($\sigma$) away the GR value (zero) occurs, from the statistical median of the posterior. 

We find that the  $\pcaOne$ parameter achieves GR exclusion at very high confidence for most of the injections. More precisely, the GR values are excluded at greater than $3\sigma$ level for most of them. The inclusion of the true value is also quantified in the same manner. The true values of $\pcaOne$ are computed using the same linear combinations that are used to compute the new posteriors and we find the true values to fall within their 1$\sigma$ bound for most of the cases. 

To understand the features in the Table~\ref{tab:nongr-pca}, it is important to recall that the detectability of a GR violation, among other things, is expected to depend strongly on (a) the strength of the deviation injected, (b) signal to noise ratio of the event, and (c) the masses of the binary constituents. We should be able to detect a GR deviation better when the strength of violation is greater and the signal is louder. Regarding the effect of mass, the ability to detect a violation will depend on whether or not the late-time dynamics of the system occurs in the most sensitive region of the noise PSD, as this will have a pronounced effect on the parameter estimation.

From Table~\ref{tab:nongr-pca}, one would notice that the most confident detections of the GR violations (simulations 1 and 3) are also accompanied by a precise quantification of the true value of the injection. Of these, simulation 3 also has the highest SNR among all the injections, with an optimal network SNR of 54. 
However, the second loudest signal, simulation 6,  which has an optimal network SNR of 27, leads to a non-GR detection only at $3.8\sigma$, compared to $13.8\sigma$ of simulation 3 and $10.8\sigma$ of simulation 2. Lastly, it is interesting to note that some of the lowest non-GR detections are seen to be for the lowest mass systems ($m\sim10M_{\odot}$) in the simulations, and the highest significant detections have relatively higher masses ($m\sim 30 M_{\odot})$. This should be due to the aforementioned feature where the late-time dynamics of a $30M_{\odot}$ should be happening closer to the sweet spot of the noise PSD compared to a $10M_{\odot}$ system. A  careful study would be needed to quantify this more precisely.

{The offsets of the peaks of the violin plots (Fig.~\ref{fig:violin-nongr}) from the injected values are very likely due to the choices of prior ranges for the original PN deformation parameters. We see prior railing similar to those in the top panel of Fig.~\ref{fig:GW151226-real} and such prior railings can translate into the observed offsets. A wider range of priors should resolve this problem. As our main goal here is to demonstrate the efficiency of the method to detect GR violations, which is already achieved in the present violin plots, a detailed study of this type we postpone for future work.} 

In summary, the salient feature seen here is that the leading order PCA deformation parameter can exclude GR, as well as include the true non-GR value, with a high confidence. This is a clear demonstration of the ability of PCA to detect any modifications to GR present in the signal.

\section{Conclusions and Future directions}\label{sec:conclusions}

We demonstrated that the problem of degeneracies in multi-parameter tests of GR can be cured by the use of Principal Component Analysis. Finding eigenvectors that diagonalize the covariance matrix of the likelihood provides a set of new parameters to test GR, of which only the most dominant one or two would suffice for a very accurate representation of the likelihood. These parameters, by construction, are also the best-estimated parameters. Using selected events from the first and second observing runs, we demonstrated the efficacy of this method and derived bounds on the newly constructed parameters which are linear combinations of the original post-Newtonian deformation parameters. Combining the information from five binary black hole mergers, we find that the significance of a possible deviation from GR is as low as $< 0.38\sigma$. We have also shown that the method based on PCA would also be very effective in a confident detection of a GR violation.

Application of this method to the binary black hole events during the third-observing runs is our next goal. The joint bounds from the events during O1-O3 would yield stringent bounds on possible departures from GR. A detailed study of the Bayes factors between GR and non-GR hypotheses is also planned. If GR is correct, the multi-parameter tests would yield a higher Bayes factor in favor of GR due to Occam's razor, though posteriors are broader (leading to weak constraints). As the PCA-based approach essentially captures the spirit of multi-parameter tests, one would also expect it to share this feature, but with reasonably well-constrained posteriors. However, the prior choices play a very important role here which requires a dedicated study. { The robustness of the PCA-based method to noise artifacts also is an interesting avenue for future investigation because if PCA-based parameters are less prone to noise artefacts, they can be handy to analyze several of the binary black hole events that have noise artefacts.} Lastly, advanced detectors such as Einstein Telescope, Cosmic explorer, and Laser Interferometric Space Antenna would permit independent estimation of two or more of the new PCA-based parameters thereby facilitating a more stringent test of GR. This could also be investigated in the future.

\acknowledgements
We thank C. Van Den Breock, Bala Iyer, N. V. Krishnendu, and  A. Yelikar for very useful discussions and inputs at various stages. We also thank Anuradha Gupta for their useful comments on the manuscript. K.G.A., M. S., and S.D. acknowledge the Swarnajayanti grant DST/SJF/PSA-01/2017-18 of the Department of Science and Technology, India.  K.G.A acknowledges the support of the Core Research Grant EMR/2016/005594 and MATRICS grant MTR/2020/000177 of the Science and Engineering Research Board of India. K.G.A and S.D also acknowledge support from Infosys Foundation. M.S. acknowledges the support from the National Science Foundation with grants PHY-1806630, PHY-2010970, and PHY-2110238. B.S.S. acknowledges the support of the National Science Foundation with grants PHYS-2012083. 

Thanks are due to computational support provided by LIGO Laboratory and supported by National Science Foundation Grants PHY-0757058, PHY-0823459. This material is based upon work supported by the NSF's LIGO Laboratory which is a major facility fully funded by the National Science Foundation. This research has made use of data obtained from the Gravitational Wave Open Science Center (www.gw-openscience.org), a service of LIGO Laboratory, the LIGO Scientific Collaboration, and the Virgo Collaboration. 
Virgo is funded by the French Centre National de Recherche Scientifique (CNRS), the Italian Istituto Nazionale della Fisica Nucleare (INFN), and the Dutch Nikhef, with contributions by Polish and Hungarian institutes.
This work makes use of \textsc{LALInference} \cite{lalinference},  \textsc{NumPy} \cite{vanderWalt:2011bqk}, \textsc{SciPy} \cite{Virtanen:2019joe}, \textsc{Matplotlib} \cite{Hunter:2007}, \textsc{jupyter} \cite{jupyter}, \textsc{dynesty} \cite{2019S&C....29..891H}, \textsc{bilby} \cite{bilby}, \textsc{corner} \cite{corner} and \textsc{seaborn} \cite{seaborn} software packages. 
This paper has been assigned the internal LIGO preprint number {P2100338}.

\appendix

\section{Bayesian inference of GW signals}\label{basics-Bayesian}
Let ${\mathcal H}$ be our hypothesis that the data $d$ is a sum of Gaussian noise and a GW signal of the model $h(\vec{\theta})$. Then, according to Bayes theorem, the posterior probability distribution of the model parameters, $\vec{\theta}$, can be written as,
\begin{equation}\label{BayesTheorem}
P(\vec{\theta}\,|\,{\mathcal H},d) = \frac{P(\vec{\theta}\,|\,{\mathcal H}) P(d\,|\,{\mathcal H},\vec{\theta})}{P(d\,|\,{\mathcal H})}
\end{equation}
where $P(\vec{\theta}\,|\,{\mathcal H})$ is the prior probability that quantifies the prior knowledge of $\vec{\theta},$ $P(d\,|\,{\mathcal H},\vec{\theta})$ is the \textit{likelihood} that quantifies the probability of  $d$ being the measured data when $\vec{\theta}$ is the true value of the parameter under the given signal model. In the presence of stationary noise, the functional form of the likelihood in the frequency domain may be written as,
\begin{equation}\label{BayesianLikelihood}
P(d\,|\,{\mathcal H},\vec{\theta}) \propto {\rm \exp}\left[- \frac{({\tilde d}-{\tilde h}\,|\, {\tilde d}-{\tilde h})}{2} \right].
\end{equation} 
Here, $\tilde{d}$ and ${\tilde h}$ are the data and the waveform model respectively in the frequency domain, and $(\tilde{a}\,|\,\tilde{b})$ denotes the noise weighted inner product defined as,
\begin{equation}
(\tilde{a}\,|\, \tilde{b}) =2 \,{\rm Re}\int_{f_{\rm low}}^{f_{\rm \rm high}}\frac{\tilde{a}(f)\,\tilde{b}(f)^{*}+\tilde{a}(f)^{*}\,\tilde{b}(f)}{S_n(f)}\,df,
\end{equation}
where $S_n(f)$ is the noise power spectral density of the detector in question, $f_{\rm low}$ and $f_{\rm high}$ are the lower and upper frequency cutoffs, respectively, and they depend on the detector's sensitive bandwidth. For Advanced LIGO and Advanced Virgo, we use $f_{\rm low} = 20 {\rm Hz}$ and $f_{\rm high}$ is the  ring-down frequency implied by the IMRPhenom waveforms that depends on the source properties. $P(d\,|\,{\mathcal H})$ in Eq.~(\ref{BayesTheorem}) is the Bayesian evidence for the hypothesis ${\mathcal H}$ which is equal to the likelihood marginalized over all the parameters $\theta_i$:
\begin{equation}
P(d\,|\,{\mathcal H}) = \int P(\vec{\theta}\,|\,{\mathcal H}) P(d\,|\,{\mathcal H},\vec{\theta}) d\vec{\theta}.
\end{equation}

\section{Likelihood reconstruction using PCA: Gaussian case}\label{reconstruction}

To demonstrate the process of reconstruction of the likelihood with new parameters obtained via principal component analysis, we consider a toy example below.


Let us consider a multi-dimensional likelihood, say, a six dimensional marginalized likelihood corresponding to the deformation parameters that we consider here. This multivariate Gaussian distribution reads
\begin{equation}\label{GaussianLikelihood}
{\mathcal P} (\theta^i ) \propto {\rm \exp}\left[ -\frac{1}{2} C_{jk}^{-1}\theta^j \theta^k\right],
\end{equation}
where, both indices $j,k$ are summed over from 1 to 6,  $\theta^i$ is any deformation parameter and the covariance matrix $C_{jk}$ encodes the widths of the resulting marginalized distribution. Such a Gaussian likelihood is a reasonable assumption in the presence of stationary Gaussian noise and in the limit of high SNR\cite{Finn92}.
The log likelihood reads
\begin{equation}
\ln {\cal L} \sim (C^{-1})_{jk}\theta^j \theta^k.
\end{equation}
Upon finding the eigenvectors $\theta^{\prime i}$, which diagonalize the covariance matrix to $(C)_{ij}^{\prime}$, one can re-write the log likelihood as
\begin{equation}
\ln {\cal L} \sim (C^{\prime})^{-1}_{jk}\theta^{\prime j} \theta^{\prime k}= (C^{\prime})^{-1}_{ii}\,(\theta^{\prime i})^2\simeq\sum_{i=1}^{N_{\rm max}}(C^{\prime})^{-1}_{ii}\,(\theta^{\prime i})^2,
\end{equation}
where $N_{\rm max}$ is the maximum number of eigenvectors that are retained after the PCA, which in our case is 2. In the last step, we have used the fact that the new covariance matrix is diagonal. 
\bibliography{references.bib} 
\bibliographystyle{apsrev}

\end{document}